\documentclass{PoS}

\newcommand{\bp}{{\bf{p}}}
\newcommand{\bq}{{\bf{q}}}
\newcommand{\br}{{\bf{r}}}

\newcommand{\bb}{{\bf{b}}}

\newcommand{\bC}{{\bf{C}}}
\newcommand{\bQ}{{\bf{Q}}}
\newcommand{\ket}[1]{| {#1} \rangle}

\def\lsim{\mathrel{\rlap{\lower4pt\hbox{\hskip1pt$\sim$}}
    \raise1pt\hbox{$<$}}}         
\def\gsim{\mathrel{\rlap{\lower4pt\hbox{\hskip1pt$\sim$}}
    \raise1pt\hbox{$>$}}}         

\title{Unitarity cutting rules for hard processes on nuclei}

\ShortTitle{Unitarity cutting rules for hard processes on nuclei}

\author{\speaker{Wolfgang Sch\"afer}%
        \thanks{Partially supported by the Polish Ministry for Science and Higher Education
under contract 1916/B/H03/2008/34.}\\
       Institute of Nuclear Physics PAN, Krak\'ow, Poland\\
       E-mail: \email{Wolfgang.Schafer@ifj.edu.pl}}

\abstract{Heavy nuclei introduce a new scale into the pQCD 
description of hard processes on nuclei, the saturation scale,
and the familiar linear $k_\perp$--factorization breaks down.
It is replaced by a new concept, the nonlinear 
$k_\perp$--factorization. 
Here we give a brief overview how topological cross sections
for hard processes on nuclei are obtained from 
nonlinear $k_\perp$--factorization.}

\FullConference{High-pT Physics at LHC - Tokaj'08 \\
          16-19 March 2008\\
          Tokaj, Hungary}

\begin{document}

\section{Introduction}
In the realm of small--$x$ physics, hard processes are adequatly 
described within (linear) $k_\perp$--factorization. The major ingredient
is the unintegrated gluon density of a nucleon \cite{Antoni_et_al}.
Heavy nuclei bring in a new scale, the saturation scale $Q_A^2(x)$
which grows with the opacity -- or size -- of the nucleus \cite{Mueller}.
From a different point of view, multiple gluon exchanges between, say,
a projectile color dipole and the target nucleus are enhanced  
by the large size of a nucleus. 
Heavy nuclei then provide us with an opportunity to study the 
physics of a regime of strong absorption/rescattering corrections
in a fairly systematic fashion, where we only need to account for 
(and ``resum'') those contributions in perturbation theory which 
grow with the size of the target.

An important issue is then what will be the fate of linear 
$k_\perp$--factorization in a nuclear environment, or more generally
in a regime of a large saturation scale.
It turned out, illustrated on the example of dijets in deep inelastic scattering 
\cite{DIS_Dijets}, 
that linear  $k_\perp$--factorization is broken and must be replaced by a new
concept, called non--linear $k_\perp$--factorization. The latter 
emerges as a generic feature of the pQCD approach to hard processes in a
nuclear environment, where hard cross sections turn out to be nonlinear
functionals of a properly defined nuclear unintegrated glue
\cite{DIS_Dijets,Nonuniversality,Single_jets,Quark_Gluon,Gluon_Gluon,
Cutting_Rules,NS_LPM}, which we will
discuss further below.

Let us stress that a statement of factorisation --or the breaking thereof-- 
is in fact one about the relations between different observables. 
For example, while in the familiar linear $k_\perp$ factorisation the spectrum
of dijets in the current fragmentation region simply maps out the transverse
momentum dependence of the same unintegrated gluon distribution which enters
the inclusive DIS structure function \cite{Dijets_Nucleon}, there is no such simple relation between
dijets and inclusive DIS on a nuclear target. 

Explicit quadratures for all cases of interest can be found in our series
of papers quoted above. Other approaches to the problem of factorization
in a saturation regime are found in the reviews \cite{CGC}.
It is quite remarkable that the formalism of nonlinear $k_\perp$--factorization
gives in a surprisingly straightforward manner access not only
to fully inclusive single-- and dijet cross sections, but also to topological
cross sections \cite{Cutting_Rules}.

Return once more to DIS: after multiple gluon exchanges between the $q \bar q$
color dipole and the nucleus, the nuclear debris will be left in a state with 
multiple color excited nucleons. 
Cross sections for final states with a fixed 
number of cut pomerons (or color excited or ``wounded'' nucleons) are
called topological cross sections. It is customary to describe 
topological cross sections in a language of unitarity cuts through
multipomeron exchange diagrams \cite{AGK}. 
In an obvious manner, color excited 
nucleons in the final state give a clear--cut definition of a cut pomeron.

Topological cross sections carry useful information on the correlation between 
forward or midrapidity
jet/dijet production and multiproduction in the nuclear fragmentation region.
They are also closely related to the important concept of centrality of a 
collision.

We will now turn to a brief review of the formalism of nonlinear 
$k_\perp$--factorisation, paying close attention to the derivation of 
topological cross section.

\section{Dijet production as excitation of beam partons $a \to bc$}
  
\begin{figure}
\begin{center}
\includegraphics[width=\textwidth]{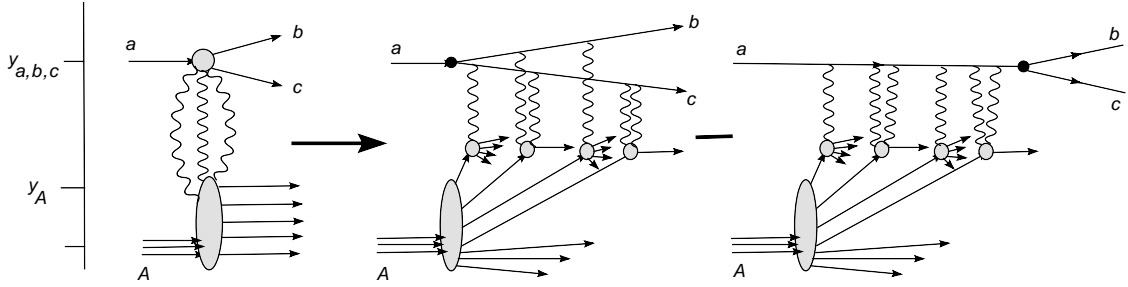}
\caption{A typical contribution to the inelastic transition $a A \to bc X$
with multiple color excitations of the nucleus. The amplitude receives 
contributions from processes with interactions before and after 
the virtual decay, which interfere destructively. 
The (pseudo)rapidities of partons $a,b,c$ must be larger, or of the order of
the nuclear boundary condition rapidity $\eta_A=\log{1\over x_A}$, 
where $x_A$ is defined in the text.}
\label{fig1}
\end{center}
\end{figure}

The most intriguing phenomena are expected in a situation where 
a hard production process is coherent over the whole longitudinal extent of
the nucleus, and
the target nucleons contribute to the process in a collective manner.
This requires a coherency condition to be fulfilled (see Fig\ref{fig1}),
which at high energies is typically the case over large areas of phase space.
For example in DIS we demand, that $x \lsim x_A =1/2R_A m_p \approx 0.1 \cdot A^{-1/3}$, 
where $R_A$ is the nuclear radius, and $m_p$ the proton mass.

In the general case we deal with the breakup of a beam parton $a$ into its 
two--body $bc$ Fock--component.
The calculation is best done in the framework of light--cone wave functions,
and in impact parameter space. Indeed the fast partons move along straight--line
trajectories, and their impact parameters are conserved during the interaction
with the target. It is only in the $a \to bc$ quantum--transition, that the
impact parameter will be `shared' according to $\bb_a = z_b \bb_b + z_c \bb_c$,
ensuring conservation of angular momentum. Here $\bb_i$ denotes the impact 
parameter of parton $i$, and $z_{b,c}$ are the light--cone momentum fractions
of $b,c$.
 
Now it would be a daunting task to calculate the amplitude for the $a \to bc$
transition with $k$ color excited nucleons in the final state, a tensor with,
among others, $k$ adjoint color indices. An elegant solution to the
multichannel intranuclear evolution problem relies on techniques developed by
Zakharov \cite{Slava}. Namely, one should turn to the $bc$ density 
matrix, and first average over the target states. The $S$--matrix of a parton
in the complex conjugate amplitude can be viewed as $S$ matrix of an antiparton,
and one ends up with an intranuclear evolution problem for a multi--(2,3,4-)parton 
system in an overall color singlet state. 

\subsection{Master formula for dijets}

We can then derive the following master formula for the differential cross
section of the $a \to bc$ process \cite{Single_jets} with $k$ color--excited
nucleons in the final state \cite{Cutting_Rules}:

\begin{eqnarray}
&&{d \sigma^{(k)} (a \to bc) \over dz_b d^2\bp_b d^2\bp_c} = \int 
{d^2\bb_b d^2\bb_c d^2\bb'_b d^2\bb'_c \over (2 \pi)^4} \exp[-i\bp_b(\bb_b-\bb_b')
- i \bp_c (\bb_c - \bb_c')] 
\nonumber \\
&&\psi_{a \to bc}(z_b , \bb_b - \bb_c) \psi_{a\to bc}^*(z_b,\bb'_b - \bb'_c) 
\nonumber \\
&&\Big\{ 
S^{(4,k)}_{\bar{b}\bar{c}cb}(\bb'_b,\bb'_c,\bb_b,\bb_c)
+S^{(2,k)}_{\bar{a}a}(\bb',\bb)
-S^{(3,k)}_{\bar{b}\bar{c}a}(\bb,\bb'_b,\bb'_c)
-S^{(3,k)}_{\bar{a}bc}(\bb',\bb_b,\bb_c)
\Big\}
\end{eqnarray}

Here $\bp_{b,c}$ are the transverse momenta of partons $b,c$, $\psi_{a \to bc}$
is the light--cone wave function for the transition $a \to bc$. 
The index $k$ reminds us of our restriction on the final state, namely it must
contain $k$ color--excited nucleons.
If we sum over all final states, the multiparton $S$--matrices can be 
evaluated using Glauber--Gribov theory. 
The building block of the multiple scattering expansion is the color dipole (CD)
cross section operator 
$\hat{\Sigma}^{(4)}(\bC)$ 
for the interaction of the $\bar b \bar c bc$--
system with a free nucleon. Here $\bC$ is a collective label for the relevant impact parameters.
$\hat{\Sigma}^{(4)}(\bC)$ 
is a matrix in the space of possible color--singlet states
(within a chosen coupling--scheme), $\ket{R \bar R} = \ket{(bc)_R (\bar b \bar c)_{\bar R}}$.
It has been obtained for all the cases of practical interest:
\begin{itemize}
\item DIS: $\gamma^* \to q\bar{q} \quad \Longrightarrow \quad \underbrace{1}_{1} + 
\underbrace{8}_{N_c^2}$
\item Open charm: $g \to c \bar{c} \quad \Longrightarrow \quad 
\underbrace{1}_{1 (N_c-{\textrm{suppressed}})}
+ \underbrace{8}_{N_c^2}$
\item Forward dijets: $q \to qg \quad \Longrightarrow \quad \underbrace{3}_{N_c} 
+ \underbrace{6 + 15}_{N_c \times N_c^2}$
\item Central dijets: $g \to gg \quad \Longrightarrow \quad \underbrace{1}_{1 (N_c-
{\textrm{suppressed}})} 
+ \underbrace{8_A + 8_S}_{N_c^2} 
+ \underbrace{10 + \overline{10} + 27 + R_7}_{N_c^2 \times N_c^2}$
.
\end{itemize} 
The color algebra was performed for $SU(N_c)$, we labelled the pertinent representations
mostly by their $SU(3)$ dimensions. For concrete applications a large--$N_c$ expansion
is helpful, and we also indicated the sizes of representations at large $N_c$. There
emerges a systematics which leads to a notion of universality classes of observables.

The crucial step is now to decompose the free--nucleon CD cross section operator into
a color rotation/excitation which represents the cut Pomeron, and the elastic part, which
corresponds to the color singlet exchange two--gluon exchange with a nucleon in the amplitude,
and represents the uncut Pomeron:

\begin{eqnarray}
\hat{\Sigma}^{(4)}(\bC)  = 
\underbrace{\hat{\Sigma}^{(4)}_{ex}(\bC)}_{\textrm{color rotation/excitation}} 
+ \underbrace{\hat{\Sigma}^{(4)}_{el}(\bC)}_{\textrm{color diagonal}}
\end{eqnarray}

It is important to realize, that the cut and uncut Pomeron parts of 
$\hat{\Sigma}^{(4)}(\bC)$ separately are infrared sensitive, and depend 
explicitly on a nonperturbative parameter, the CD cross section for a large
dipole $\sigma_0$.

Returning to the nuclear problem, we obtain, for example for the four--parton
$S$--matrix from Glauber--Gribov theory:

\begin{eqnarray}
S^{(4)}_{\bar{b}\bar{c}cb}(\bC)
= \sum_k
S^{(4,k)}_{\bar{b}\bar{c}cb}(\bC)
=
\exp \Big[-{1 \over 2} T_A(\bb) \,
\big( \hat{\Sigma}^{(4)}_{ex}(\bC) +\hat{\Sigma}^{(4)}_{el}(\bC) \big) \Big] 
\, ,
\label{S4}
\end{eqnarray}
where $T_A(\bb)$ is the well--known nuclear thickness function.
Now, the sought--for multiparton $S$--matrix for the final state with
$k$ color excited nucleons can be obtained from the $k$-th order term
of the expansion of (\ref{S4}):
\begin{eqnarray}
S^{(4,k)}_{\bar{b}\bar{c}cb}(\bC)
&&= (-1)^k \int_0^1 d\beta_k \dots \int_0^1 d\beta_1 \, 
G_0(1-\beta_k, \bC) 
\hat{\Gamma}_{ex}(\bC) 
G_0(\beta_k - \beta_{k-1},\bC) \dots
\hat{\Gamma}_{ex}(\bC) 
G_0(\beta_1,\bC) 
\nonumber \\
&& + \delta_{k,0} G_0(1, \bC) \, ,
\end{eqnarray}
where
\begin{equation}
G_0(\beta,\bC) = \theta(\beta)
\exp \Big[-\beta {1 \over 2} T_A(\bb) \hat{\Sigma}^{(4)}_{el}(\bC) \Big]  \, , \, 
\hat{\Gamma}_{ex}(\bC) = {1 \over 2} T_A(\bb) \hat{\Sigma}^{(4)}_{ex}(\bC) 
\, .
\end{equation}

Here the parameter $\beta$ has the meaning of a dimensionless depth inside the 
nucleus. Notice that the nested $\beta$--integration arises due to the fact that
$  \hat{\Sigma}^{(4)}_{ex} $ and $\hat{\Sigma}^{(4)}_{el}$ do not commute.
This underlines the fact that the nucleus cannot be treated in terms of a classical
field of the target as a whole.
Furthermore, an expansion of the exponential in $G_0$ would give rise to the familiar
alternating sign expansion of uncut multipomeron absorptive corrections. Ultimately,
the contributions from $G_0$'s can be regrouped into effective coherent distortions
of the lightcone wave--function for the $a \to bc$ transition. This is a typical
manifestation of the physics of large coherence lengths.

\section{Nuclear unintegrated glue and its properties}

After we have established all ingredients of the calculation in impact 
parameter space, let us move on to the momentum space formulation.
Here the central quantity is the nuclear unintegrated glue. We remind the reader,
that in the coherent breakup of pions into dijets, the diffractive final state 
consist of a back--to--back dijet in which the large transverse momenta 
of jets are taken from gluons exchanged with the target nucleons \cite{NSS}. Thus 
the diffractive amplitude, or the $S$--matrix of a $q\bar q$--dipole,
serves as a good definition of the collective
nuclear unintegrated glue:
\begin{eqnarray}
\Phi(\bb,x,\bp) = \int{d^2 \br \over (2 \pi)^2} S_{q\bar q} (\bb,x,\br) \exp[-i \bp \br] 
= \exp[-\nu_A(\bb)] \delta^{(2)}(\bp) + \phi(\bb,x,\bp)  
\, ,
\end{eqnarray}
Here $\nu_A(\bb) = \sigma_0(x) T_A(\bb)/2$ is the nuclear opacity.
At the boundary value $x=x_A$, one can derive a useful representation
of the collective nuclear unintegrated glue in terms of the free-nucleon
unintegrated gluon structure function (we use a notation
$f(x,\bp) \propto \bp^{-4} \partial G(x,\bp^2)/ \partial \log(\bp^2)$):
\begin{eqnarray}
\phi(\bb, x_A, \bp) =  \sum_k 
w_k \big(\nu_A(\bb) \big) f^{(k)} (\bp) \, , \,
 f^{(k)} (\bp) = \displaystyle \int \big[ \prod^k d^2 \bp_i 
f(\bp_i) \big] \delta^{(2)}( \bp- \sum \bp_i)  \, , 
\label{Nuc_glue}
\end{eqnarray}
and
\begin{equation}
w_k\big( \nu_A(\bb) \big) = {\nu^k_A(\bb) \over k !} \, \exp[-\nu_A(\bb)] \, .
\end{equation}
Clearly, we observe here once more the equivalence between the 
1975--parton fusion description of nuclear shadowing \cite{NZ75} and the 
unitarization of the color dipole--nucleus interaction \cite{NZ91}. 

We come to the first of unitarity cutting rules in momentum space.
Firstly, there holds a proportionality of the nuclear unintegrated glue
and the quasielastic inclusive single quark cross section:
\begin{equation}
{d \sigma(q A \to qX) \over d^2 \bb d^2 \bp} = \phi(\bb,x_A,\bp) \, ,
\end{equation}
and secondly, the $k$--th order term in the expansion (\ref{Nuc_glue}) 
is \emph{precisely} the topological cross section for the quark--nucleus 
scattering with $k$ color excited nucleons in the final state at $x=x_A$:
\begin{equation}
{d \sigma^{(k)} (qA \to qX) \over d^2 \bb d^2 \bp} = 
w_k \big(\nu_A(\bb) \big) f^{(k)} (\bp)  \, ,
\end{equation}

\subsection{Nuclear unintegrated glue: salient features}
\begin{figure}
\begin{center}
\includegraphics[width=.5\textwidth,angle=270]{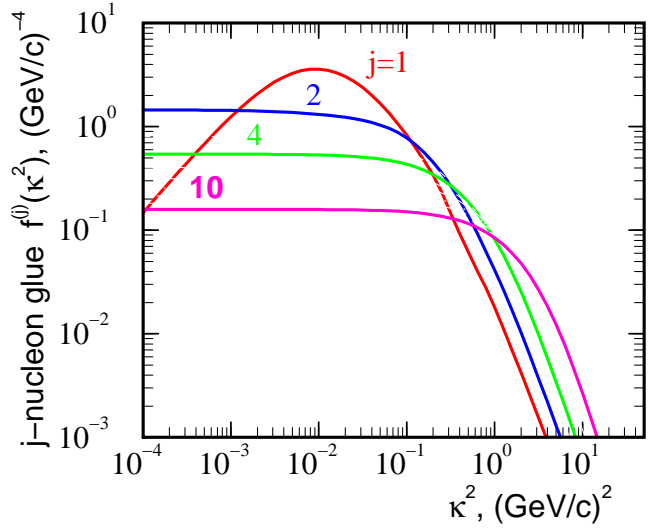}
\caption{Multiple convolution of the free nucleon unintegrated glue 
(including gluon propagators) at $x=0.01$. Observe the emergence of the plateau
in higher convolutions.}
\label{fig2}
\end{center}
\end{figure}

Before proceeding to the more complex dijet observables, let us
collect a few salient features of the collective nuclear glue,
which follow from the representation (\ref{Nuc_glue}) \cite{NSS}.
Firstly, for soft gluon momenta (small $\bp$), the collective glue
develops a plateau of the form (see Fig.\ref{fig2})
\begin{eqnarray}
\phi(\bb,x_A.\bp) \sim {1 \over \pi} 
{Q_A^2(\bb,x_A) \over (\bp^2 + Q_A^2(\bb,x_A) )^2 } \, ,
\end{eqnarray}
where the width of the plateau is the saturation scale
$ Q_A^2(\bb,x_A) \sim {4 \pi^2 \over N_c} \alpha_S(Q_A^2) G(x,Q_A^2) T_A(\bb)$,
and $G(x,Q^2)$ is the collinear gluon structure function of a nucleon.

For the hard tail $\bp^2 \gsim Q_A^2$, one obtains a Cronin--type antishadowing of  
the glue per bound nucleon:
\begin{equation}
{\phi(\bb,x_A,\bp) \over \nu_A(\bb)} = f(x_A,\bp) \Big[ 1 + {\gamma^2 \over 2}
\cdot {\alpha_S(\bp^2) G(x_A,\bp^2) \over \alpha_S(Q_A^2) G(x_A,Q_A^2) } \cdot 
{Q^2_A(\bb,x_A) \over \bp^2 }\Big] 
\, ,
\end{equation}
where $\gamma \gsim 2$ is the exponent of the hard tail of $f(x,\bp) \propto \alpha_S(\bp^2)/ \bp^{2
\gamma}$. A remarkable fact about these large and small $\bp^2$ behaviours is 
that they are predicted without any soft parameter.
Furthermore, regarding its small--$x$ dependence, it can be shown \cite{NS_LPM}
that $\phi(\bb,x,\bp)$ fulfills the Balitsky--Kovchegov \cite{BK} equation.

\section{Topological cross sections for dijet processes}

Let us finally demonstrate our cutting rules on the more involved example 
of dijets in the current fragmentation region in DIS. 
The nonlinear $k_\perp$--factorization expression for the differential
cross section, reads (we omit the possible diffractive final state):
\begin{eqnarray}
{(2 \pi)^2 d \sigma_A(\gamma^* \to q(\bp) \bar q (\bQ-\bp)) \over d^2\bb dz d^2\bp d^2\bQ}
&&= {1 \over 2} T_A(\bb) \int_0^1 d \beta \int d^2 \bq_1 d^2 \bq f(\bq)
\Big|\psi(\beta,z,\bp - \bq_1) - \psi(\beta, \bp - \bq_1 -\bq) \Big|^2
\nonumber \\
&& \times
\Phi(1-\beta, \bb,\bQ - \bq_1 - \bq ) \Phi(1-\beta, \bb,\bq_1)
\, .
\label{Nlin}
\end{eqnarray}
Here we omitted to show the argument $x_A$. $\Phi(1-\beta,\bb,\bq)$
is the nuclear unintegrated glue taken for the opacity $(1-\beta) \nu_A(\bb)$,
and the light--cone wave function of the virtual photon coherently
distorted over a slice $\beta$ of the nucleus is given by
a convolution over transverse momenta  
$\psi(\beta,z,\bp) = ( \Phi(\beta) \otimes \psi ) (\bp)$.
Now, the unitarity cutting rule is easily stated: to obtain the partial
cross section with $k$ color excited nucleons in the final state, substitute
in the nonlinear $k_\perp$--factorization formula \ref{Nlin}
(we skip all arguments except transverse momenta):
\begin{equation}
\Phi(\bp_1) \Phi(\bp_2) \to \sum_{i,j} \delta(k-1-i-j) w_i \, w_j f^{(i)}(\bp_1) 
f^{(j)}(\bp_2) \, . 
\end{equation}
A few comments: we witness in equation \ref{Nlin} the emergence of 
\emph{two types of cut Pomerons}.
One, of the color excitation type, which is related to the transition of the 
$q \bar q$ dipole into the color octet state, and another one of the color rotation type
-- once in the color octet, quark and antiquark scatter independently in 
the remaining slice of the target. Any regeneration of the initial color singlet state
is large--$N_c$ suppressed \cite{DIS_Dijets}, yet of course explicitly calculable.
The two types of cut Pomerons are but a technical manifestation 
of the multichannel property of the intranuclear evolution problem. 
In fact this property makes much of the standard Glauber--AGK lore
from old--fashioned hadronic models inapplicable to pQCD, see for 
example \cite{Cutting_Rules,HSQCD}.

Let us turn to an interesting feature of topological cross sections for single particle 
spectra. When integrating the expression for the $k$--cut Pomeron dijet 
cross section over the transverse momentum of one of the jets, we realise,
that the factors $w_j$ of the spectator parton do not cancel out. Notice
that the cancellation of spectator interactions is an important ingredient of
the pQCD factorization theorems -- indeed it relies on the fact that we sum over 
all final states. Interestingly, we can recover the quark contribution
through a peculiar resummation over backward multiplicities \cite{Cutting_Rules}. 

\section{Summary/Outlook}
At small--$x$, strongly absorbing targets, like heavy nuclei, introduce
a new scale into the pQCD description of hard processes, the saturation scale.
In such a regime, the conventional linear $k_\perp$ factorization breaks down, 
and is replaced by the new concept of nonlinear $k_\perp$ factorization.
We have demonstrated how the nonlinear $k_\perp$ factorization formulas
for single-- and dijet processes give rise to the corresponding expressions
for topological cross sections. First steps for phenomenological apllications
to DIS structure functions have been taken \cite{HSQCD}. 
Of particular interest will be the analysis of quenching of forward jets
in $pA$ and $\gamma^* A$ processes. Obviously there must be an extra flow 
of energy from the forward region to the nuclear hemisphere, depending
on the added activity due to color excited nucleons.
 
\vspace{1cm}
\noindent
{\bf{Acknowledgements:}}
It is a pleasure to thank the organizers for the hospitality at this 
informative workshop, and Kolya Nikolaev, Slava Zakharov and Volodya Zoller
for collaboration on the work presented here.


\begin{thebibliography}{99}

\bibitem{Antoni_et_al}
see for example the contributions by A. Szczurek, M. \L uszczak and G. \'Slipek
in these proceedings, and references therein.

\bibitem{Mueller}
  A.~H.~Mueller,
  Nucl.\ Phys.\  B {\bf 335} (1990) 115; Nucl.\ Phys.\  B {\bf 558} (1999) 285.

\bibitem{DIS_Dijets}
  N.~N.~Nikolaev, W.~Sch\"afer, B.~G.~Zakharov and V.~R.~Zoller,
  J.\ Exp.\ Theor.\ Phys.\  {\bf 97} (2003) 441
  [Zh.\ Eksp.\ Teor.\ Fiz.\  {\bf 124} (2003) 491]

\bibitem{Nonuniversality}
  N.~N.~Nikolaev, W.~Sch\"afer and B.~G.~Zakharov,
  Phys.\ Rev.\ Lett.\  {\bf 95} (2005) 221803
  [arXiv:hep-ph/0502018].

\bibitem{Single_jets}
  N.~N.~Nikolaev and W.~Sch\"afer,
  Phys.\ Rev.\  D {\bf 71} (2005) 014023
  [arXiv:hep-ph/0411365].

\bibitem{Quark_Gluon}
  N.~N.~Nikolaev, W.~Sch\"afer, B.~G.~Zakharov and V.~R.~Zoller,
  Phys.\ Rev.\  D {\bf 72} (2005) 034033
  [arXiv:hep-ph/0504057].

\bibitem{Gluon_Gluon}
  N.~N.~Nikolaev, W.~Sch\"afer and B.~G.~Zakharov,
  Phys.\ Rev.\  D {\bf 72} (2005) 114018
  [arXiv:hep-ph/0508310].

\bibitem{Cutting_Rules}
  N.~N.~Nikolaev and W.~Sch\"afer,
  Phys.\ Rev.\  D {\bf 74} (2006) 074021.


\bibitem{NS_LPM}
  N.~N.~Nikolaev and W.~Sch\"afer,
  Phys.\ Rev.\  D {\bf 74} (2006) 014023
  [arXiv:hep-ph/0604117].

\bibitem{Dijets_Nucleon}
  A.~Szczurek, N.~N.~Nikolaev, W.~Sch\"afer and J.~Speth,
  Phys.\ Lett.\  B {\bf 500} (2001) 254
  [arXiv:hep-ph/0011281].


\bibitem{CGC}
  J.~Jalilian-Marian and Y.~V.~Kovchegov,
  Prog.\ Part.\ Nucl.\ Phys.\  {\bf 56} (2006) 104
  [arXiv:hep-ph/0505052];
  F.~Gelis, T.~Lappi and R.~Venugopalan,
  Int.\ J.\ Mod.\ Phys.\  E {\bf 16} (2007) 2595
  [arXiv:0708.0047 [hep-ph]].

\bibitem{AGK}
  V.~A.~Abramovsky, V.~N.~Gribov and O.~V.~Kancheli,
  Yad.\ Fiz.\  {\bf 18} (1973) 595
  [Sov.\ J.\ Nucl.\ Phys.\  {\bf 18} (1974) 308.

\bibitem{Slava}
  B.~G.~Zakharov,
  Nucl.\ Phys.\ Proc.\ Suppl.\  {\bf 146} (2005) 151
  [arXiv:hep-ph/0412117] and references therein.

\bibitem{NSS}
  N.~N.~Nikolaev, W.~Sch\"afer and G.~Schwiete,
  Phys.\ Rev.\  D {\bf 63} (2001) 014020
  [arXiv:hep-ph/0009038].

\bibitem{NZ75}
  N.~N.~Nikolaev and V.~I.~Zakharov,
  Phys.\ Lett.\  B {\bf 55} (1975) 397;
  Sov.\ J.\ Nucl.\ Phys.\  {\bf 21} (1975) 227
  [Yad.\ Fiz.\  {\bf 21} (1975) 434].

\bibitem{NZ91}
  N.~N.~Nikolaev and B.~G.~Zakharov,
  Z.\ Phys.\  C {\bf 49} (1991) 607.

\bibitem{BK}
  I.~Balitsky,
  Nucl.\ Phys.\  B {\bf 463} (1996) 99
  [arXiv:hep-ph/9509348];
  Y.~V.~Kovchegov,
  Phys.\ Rev.\  D {\bf 60} (1999) 034008
  [arXiv:hep-ph/9901281].

\bibitem{HSQCD}
  W.~Sch\"afer,
  ``Unitarity Cutting Rules for Hard Processes on Nuclear Targets,''
  arXiv:0809.0861 [hep-ph].

\end{thebibliography}
\end{document}